# Towards standardisation of average grain size measurement of additively manufactured microstructures using EBSD

Vivian Tong[1], Hannah Zhang[1], Jacopo Del Gaudio[1,2], Ken Mingard[1], Ali Gholinia[2]

1. National Physical Laboratory, Teddington, United Kingdom
2. Henry Royce Institute, Department of Materials, University of Manchester, Manchester, M13 9PL, United Kingdom

## 1 Introduction

Additively manufactured (AM) alloys have heterogeneous microstructures with broad grain size distributions and highly anisotropic and/or non-convex grain shapes. AM components can have complex geometries and porosity which may affect the local microstructure. Currently there is no electron backscatter diffraction (EBSD)-based grain size measurement standard suitable for typical AM materials.

An interlaboratory comparison study was conducted to find out what grain size metrics and summary statistics are currently used to describe average grain size. Participants were asked to measure and report the average grain size from the same EBSD map dataset. Detailed reports have been published in Reference [1]. Based on these results, we have tested and propose recommendations for a new standard for measuring average grain size in AM materials. The present work demonstrates the suitability and limitations of the proposal across several different Ni and Al AM components.

### 1.1 Grain size standards

A major standardisation need of grain size measurement is for patent claims and alloy grade specifications. The measured grain size in this context is to distinguish between (dis)similar structures. Three general standards exist for grain size measurement of metallic materials: ASTM E112 [2], describes intercept and planimetric methods to measure average grain size from micrograph images; ASTM E2627 [3] and ISO 13067 [4] both describe electron backscatter diffraction (EBSD)-based grain size measurements, but specify different requirements. Other material-specific standards for grain size measurement also exist, such as ISO 4499 for WC grain size measurement in hardmetals.

ASTM E112 has been cited within the 'Claims' section of 242 patents between 1970 and 2024 [5], but neither ASTM E2627 nor ISO 13067 have been cited in the 'Claims' section of any patents since their first publications in 2013 and 2011 respectively [6].

Quantitative grain boundary identification and full sectional grain size distributions are available from EBSD orientation maps. For additively manufactured (AM) alloys, which typically have complex microstructures with broad grain size distributions, EBSD may offer a more complete description of the microstructure compared to planimetric or intercept-based methods which only measure the average grain size. However, EBSD-based average grain size measurements are extremely sensitive to operator choices during the experimental, data clean-up, grain reconstruction, and statistical analysis steps, which may limit its usefulness in a patent claim.

## 1.2 Grain size metrics

The simplest grain size metric available from an EBSD grain map is grain area. This metric makes no assumptions about grain shape, but area units ($m^2$) are less intuitive than length units, so the equivalent circle (CEq) diameter is often reported instead. This approximation of the grain sectional diameter in equiaxed microstructures, but can be extremely misleading if the grains are elongated, e.g. along the build direction. Ellipse fitting can describe the elongation, but the fitted ellipse shape is sensitive to the choice of best-fit algorithm, especially for highly non-convex grains. Feret diameters along various directions, which is the minimum distance between two parallel lines containing the grain, can be used instead. These values are lengths that can be measured directly from the EBSD grain map. The maximum Feret diameter (MFD) is most commonly reported. The minimum Feret diameter, as well as Feret diameters along primary sample axes, are also used. The ASTM number G is specified in ASTM grain size standards to represent average grain size. It is convenient for relating materials properties because it scales with most grain-size-dependent material properties [7], but it is unintuitive to relate to microstructural features on an EBSD map.

## 1.3 Grain size statistics

The simplest statistic to summarise a grain size distribution is the (implicitly number-weighted, arithmetic) **mean**. The **number-weighted** mean tends to underestimate the grain size, since small grains and misindexed points are weighted equally to large grains. The **area-weighted** mean is a better option as it weights each grain's contribution to the size distribution by its sectional area. Use of the **arithmetic** mean implies a normal distribution shape, whereas the **geometric** mean value implies a log-normal distribution, which is both empirically a better match of the grain size distribution shape, and consistent with normal grain growth and recrystallisation theories [8], [9]. Alternatively, **quantile values** such as the 10$^{th}$, 50$^{th}$, and 90$^{th}$ percentiles (written as D10, D50, and D90 in ISO 13067) do not assume anything about the underlying distribution shape, and can be expressed as either number- or area-weighted values.

## 1.4 Recommended updates to existing standards

References [1], [7], [10], [11], [12] recommended several updates to both ASTM E2627 and ISO 13067, both generally and for AM materials. Items which have not already been addressed in a subsequent version of the standards are summarised below:

(1) Small grains < 10 or < 100 points per grain should not be removed as this severely biases the average grain size.

(2) The step size requirement could be relaxed. 20% of the average grain diameter (20 points per grain) is adequate; 80-500 points per grain is too much.

(3) Gaussian statistics are inappropriate as the grain size distributions are far from Gaussian. Log-normal or Weibull statistics (e.g. using a geometric instead of arithmetic mean) could be used instead. Also, EBSD measures the sectional grain size distribution, not just the average grain size; this additional information should be utilised.

(4) Obtaining EBSD maps of high enough quality (> 90 % indexed fraction) is experimentally challenging, but absolutely crucial to accurate grain size measurement. In AM microstructures, unindexed-points along low-angle or sub-grain boundaries can break up large grains (or, conversely, clean-up procedures can artificially link them).

(5) A 10° misorientation threshold angle may be more appropriate than 5° when analysing maps with highly developed sub-grain structure.

(6) The EBSD map field of view should follow the average aspect ratio of the grains.

# 2 Grain reconstruction methods and size metrics

This section aims to determine the best grain reconstruction procedure and size metric, subject to the following constraints:

(1) The procedure and results should be software-agnostic. The map post-processing procedure must be available in any of the major commercial and open-source software packages (OIM Analysis, AZtecCrystal, and MTEX), and the measured grain sizes should be consistent between them.

(2) A good grain size metric (e.g. grain area, equivalent circle diameter, minimum or maximum Feret lengths) is insensitive to operator-selected experimental or data-processing parameters, and sensitive to physical differences in the microstructure.

(3) A good grain size statistic (e.g. mean, median, $n^{th}$ percentiles) and visualisation method (e.g. grain map, histogram plot) should be easy to read (e.g. fewer numbers is better), intuitive to understand (e.g. familiar measurement units), and conservative in nature, i.e. low-quality data can be identified quickly and directly. As with the grain size metrics, it should also be insensitive to operator-selected parameters, and sensitive to physical differences.

## 2.1 Grain reconstruction

Consistency between the grain reconstruction and EBSD map clean-up methods in different software packages was tested using a small EBSD map (161 × 160 points, 1.251 μm step size), acquired using the Oxford Instruments EBSD system (AZtec 6.2) and exported in .h5oina (HDF5 binary) and .ctf (text) formats. Edge grains were not excluded because this map is very small.

Grain reconstruction was performed in the native analysis software AZtecCrystal (.h5oina), OIM Analysis 7 (.ctf import), and MATLAB 2024a running MTEX 6.0.0 (.h5oina import). Similar settings were used throughout: a single g.b. threshold angle of 10°, minimum grain size of 1 point, map edge grains included. The MTEX code is `grains=calcGrains(ebsd,'angle',10*degree);` where `ebsd` is the imported EBSD map with type `@EBSD`.

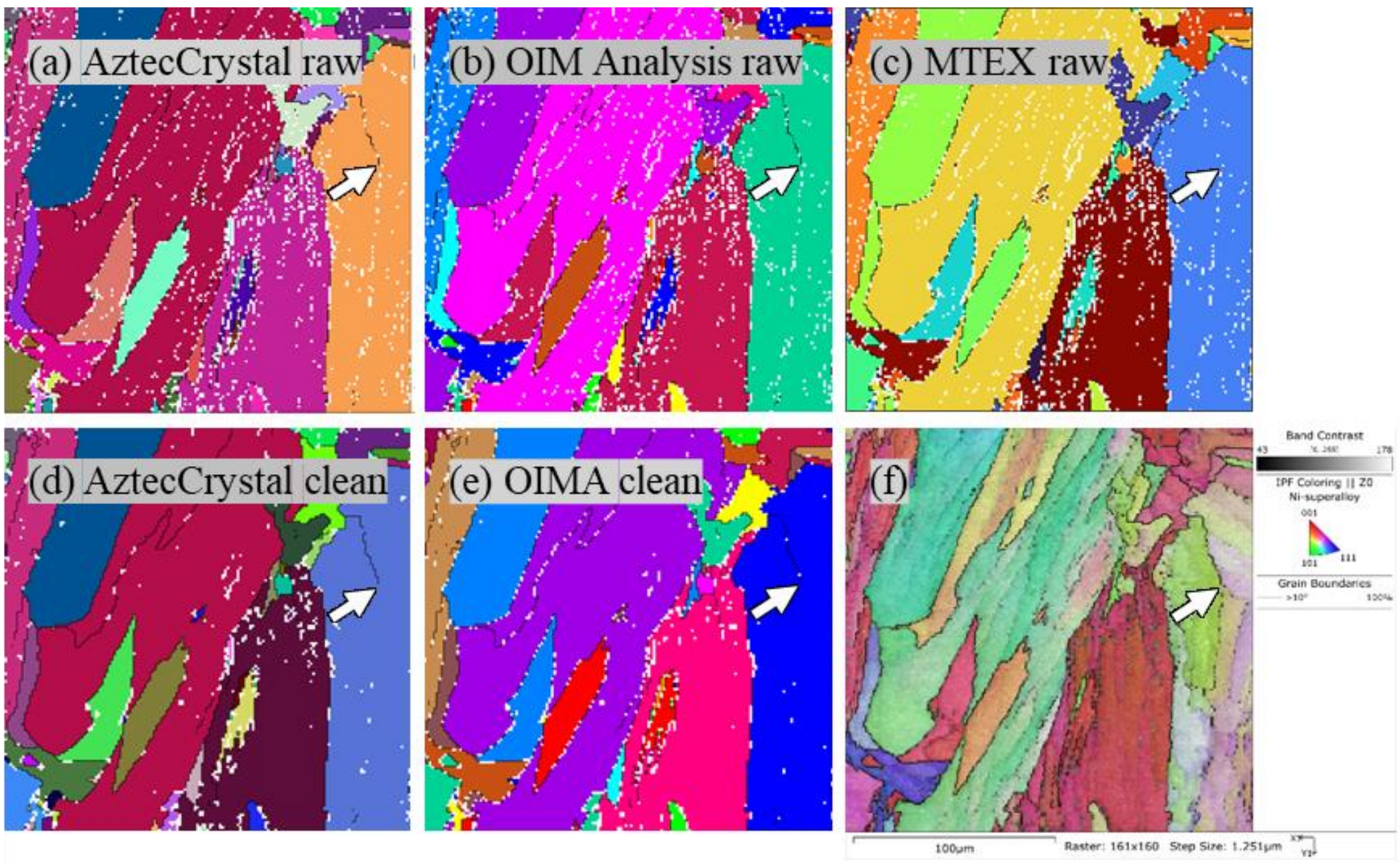

*Figure 1: EBSD data subset used to test equivalence of grain reconstruction methods. (a) – (e): Grains in random colours with grain boundaries (≥ 10°) highlighted in black. Unindexed points are white. White arrows: By default, MTEX hides incomplete boundaries whereas OIM Analysis and AZtecCrystal shows them. (e): EBSD orientation map showing band contrast and IPF-Z (out of screen) directions overlaid.*

Grain area can be computed using two methods: either multiply the EBSD map pixel area by the number of points belonging to a grain, or construct a polygon from a grain's boundaries and compute its area. Figure 2 shows the difference this makes in grains containing internal unindexed points. AZtecCrystal and OIM Analysis use the point counting method only, whereas in MTEX has both: `grains.area` uses the polygon construction method and the output is in units of $\mu m^2$, and `grains.grainSize`[1] uses the point counting method and the output is in units of pixels.

[1] The `grains.grainSize` property in MTEX 6.0.0 (released November 2024) has since been deprecated and renamed as `grains.numPixel` in MTEX 6.1.beta2 (March 2025).

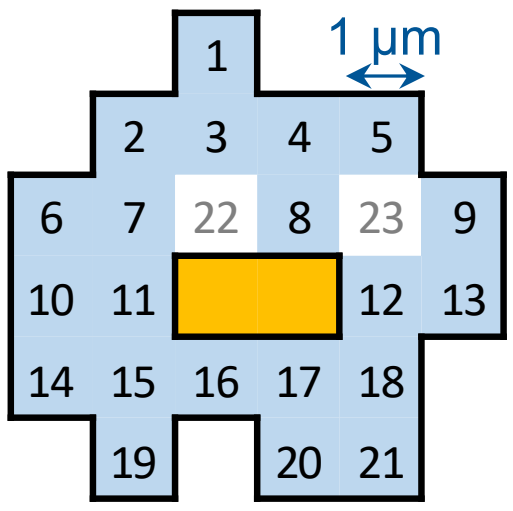

*Figure 2: Schematic showing different grain area computation algorithms. The grain of interest is blue with black boundaries. Yellow points belong to a different grain. Unindexed points are white. AZtecCrystal and OIM Analysis counts the number of blue points and computes the grain area as 21 µm². MTEX computes* `grains.area` *as 23 µm² and* `grains.grainSize` *as 21 pixels.*

## 2.2 EBSD map clean-up

No equivalent clean-up methods are available in the three software programs. Therefore, the maximum allowable clean-up routines compliant with ISO 13067 (≤ 5 % change to EBSD map) were performed in AZtecCrystal (remove wild spikes, remove zero solutions level 2) and OIM Analysis (neighbour orientation correlation level 2). AZtecCrystal's 'Autoclean' function, which for this dataset is compliant with ASTM E2627 (≤ 10 % change) but not ISO 13067, was also tested. The MTEX method [13] is different: by default, it does not clean-up the EBSD map, but assigns the map area to grains by interpolating unindexed regions. The effective 'change' to the EBSD map is therefore equal to the fraction of unindexed points (6.69 % for this data). The code to implement this is

```
grains=calcGrains(ebsd('indexed'),'angle',10*degree).
```

Grain maps (grains in random colours) from the three software programs are shown in Figure 1. The reconstructed grains are identical as long as no clean-up is performed. Figure 1 (d) and (e) show cleaned-up maps using the AZtecCrystal and OIM Analysis algorithms. There are minor differences but the overall grain structures appear qualitatively similar.

The uncertainty component from clean-up varies between datasets, but an uncertainty range can be determined from the difference between no clean-up and maximum clean-up and modelled as a Type B uncertainty with rectangular distribution. Examples will be shown later on a set of Ni-base superalloy and AlSi10Mg samples in Section 3, including two AlSi10Mg maps where the indexing rate was very low.

## 2.3 Grain size metrics

*Table 1: Grain size summary statistics of the same EBSD map using different software packages and clean-up methods using a g.b. threshold angle of 10°.*

| | | Column | 1 | 2 | 3 | 4 | 5 | 6 | 7 | 8 | 9 | 10 |
|---|---|---|---|---|---|---|---|---|---|---|---|---|
| Ni AJGQ Subset 2 | | | AztecCrystal 3.2 (native h5oina) | | | MTEX 6.0.0 (h5oina import) | | | | OIMA 7 (ctf import) | | ± Half-range (%) 20% |
| 93.31 % indexed | | | No clean | Clean[1] | "Autoclean" | No clean | | Interpolated unindexed* | | No clean | Clean[2] | |
| | | | | | | Count pixels | polygon area | Count pixels | Polygon area | | | |
| % points changed | | | 0 | 4.11 | 6.48 | 0 | | 6.69** | | 0 | 3.49 | 0% |
| Num. grains | | | 111 | 85 | 64 | 111 | 111 | 63 | 63 | 111 | 88 | 26.77% |
| Grain areas / pixels | Num-weighted | Mean | 217 | 295 | 402 | 217 | 218 | 382 | 411 | 217 | 283 | 33.04% |
| | | Geomean | 7 | 11 | 21 | 7 | 7 | 19 | 25 | 7 | 10 | 71.05% |
| | | D10 | 1 | 1 | 1 | 1 | 1 | 1 | 1 | 1 | 1 | N/A |
| | | D25 | 1 | 2 | 2 | 1 | 1 | 2 | 4 | 1 | 1 | 90.00% |
| | | D50 | 3 | 6 | 14 | 3 | 3 | 16 | 18 | 3 | 4 | 96.43% |
| | | D75 | 24 | 57 | 85 | 24 | 24 | 85 | 101 | 24 | 52 | 72.79% |
| | | D90 | 272 | 408 | 596 | 272 | 277 | 569 | 596 | 272 | 394 | 39.88% |
| | | Max | 8130 | 8476 | 8603 | 8130 | 8147 | 8138 | 8622 | 8130 | 8388 | 2.96% |
| | Area-weighted | Mean | 4412 | 4599 | 4692 | 4412 | 4420 | 4453 | 4696 | 4412 | 4600 | 3.14% |
| | | Geomean | 2629 | 2796 | 2902 | 2629 | 2633 | 2745 | 2867 | 2629 | 2820 | 4.98% |
| | | D10 | 343 | 382 | 415 | 343 | 343 | 383 | 385 | 343 | 405 | 9.69% |
| | | D25 | 2180 | 2228 | 2238 | 2180 | 2187 | 2180 | 2243 | 2180 | 2212 | 1.43% |
| | | D50 | 4422 | 4563 | 4603 | 4422 | 4458 | 4426 | 4652 | 4422 | 4570 | 2.55% |
| | | D75 | 8130 | 8476 | 8603 | 8130 | 8147 | 8138 | 8622 | 8130 | 8388 | 2.96% |
| | | D90 | 8130 | 8476 | 8603 | 8130 | 8147 | 8138 | 8622 | 8130 | 8388 | 2.96% |
| | | Max | 8130 | 8476 | 8603 | 8130 | 8147 | 8138 | 8622 | 8130 | 8388 | 2.96% |
| Max Feret diameters / µm | Num-weighted | Mean | 17 | 22 | 28 | 17 | 17 | 30 | 30 | 17 | 21 | 29.40% |
| | | Geomean | 6 | 8 | 11 | 6 | 6 | 12 | 12 | 6 | 7 | 36.49% |
| | | D10 | 2 | 2 | 2 | 2 | 2 | 2 | 2 | 2 | 2 | 0.00% |
| | | D25 | 2 | 3 | 3 | 2 | 2 | 4 | 4 | 2 | 2 | 37.50% |
| | | D50 | 4 | 6 | 9 | 4 | 4 | 11 | 11 | 4 | 5 | 54.31% |
| | | D75 | 14 | 20 | 28 | 14 | 14 | 32 | 32 | 14 | 17 | 43.78% |
| | | D90 | 47 | 54 | 83 | 47 | 47 | 83 | 83 | 47 | 52 | 29.83% |
| | | Max | 232 | 233 | 233 | 232 | 232 | 232 | 232 | 232 | 232 | 0.22% |
| | Area-weighted | Mean | 163 | 163 | 167 | 163 | 163 | 167 | 166 | 163 | 167 | 1.21% |
| | | Geomean | 139 | 141 | 145 | 139 | 139 | 146 | 144 | 139 | 144 | 2.47% |
| | | D10 | 50 | 50 | 50 | 50 | 50 | 54 | 54 | 50 | 50 | 3.93% |
| | | D25 | 118 | 119 | 119 | 119 | 119 | 119 | 119 | 118 | 118 | 0.42% |
| | | D50 | 185 | 185 | 185 | 186 | 186 | 186 | 186 | 185 | 185 | 0.27% |
| | | D75 | 232 | 233 | 233 | 232 | 232 | 232 | 232 | 232 | 232 | 0.22% |
| | | D90 | 232 | 233 | 233 | 232 | 232 | 232 | 232 | 232 | 232 | 0.22% |
| | | Max | 232 | 233 | 233 | 232 | 232 | 232 | 232 | 232 | 232 | 0.22% |

1 ISO 13067 compliant - remove wild spikes, remove zeros level 2 (i.e. min 7 neighbours)

2 ISO 13067 compliant - neighbour orientation correlation level 2

* This is the default MTEX method

** Interpolating unindexed points assigns full map area to grains; no EBSD data changed

Although the cleaned-up maps in Figure 1 (d) appear similar and have similar % changed orientations (Table 1), they produce extremely different number-weighted metrics (Table 1 Column 10). The number-weighted metrics vary by a half-range of ± 26 – 100 %. In contrast, area-weighted metrics are an order of magnitude less sensitive to clean-up method at ± 0.2 – 10 %.

The number-weighted 10$^{th}$ percentile (D10) values describe grains containing only a single point, which is related to the EBSD map step size, not the microstructure. Area-weighted D75, D90 and D100 values all describe the largest grain in the system. In this small EBSD map, the largest grain occupies about 30 % of the map area. For microstructures with extremely broad grain size distributions, quartile values (25$^{th}$, 50$^{th}$ i.e. median, and 75$^{th}$ percentiles) may be better statistical descriptors of the grain size distribution than D10/D50/D90.

Table 1 Columns 1, 4 and 8 show the grain size measurements with no data cleaning. The reconstructed grain areas are numerically identical for all three methods. The MFDs are numerically identical in OIM Analysis and AZtecCrystal, and negligibly different in MTEX (± << 1 pixel length).

The effect of grain area computation method on grain size is shown in Table 1 (Columns 3 vs 4, and Columns 5 vs 6). Its uncertainty contribution to grain area is similar in size to the effect of data clean-up, but its effect on MFD is negligible, as it only affects the area-weighting, not the MFD length measurements.

## 2.4 Excluded grains

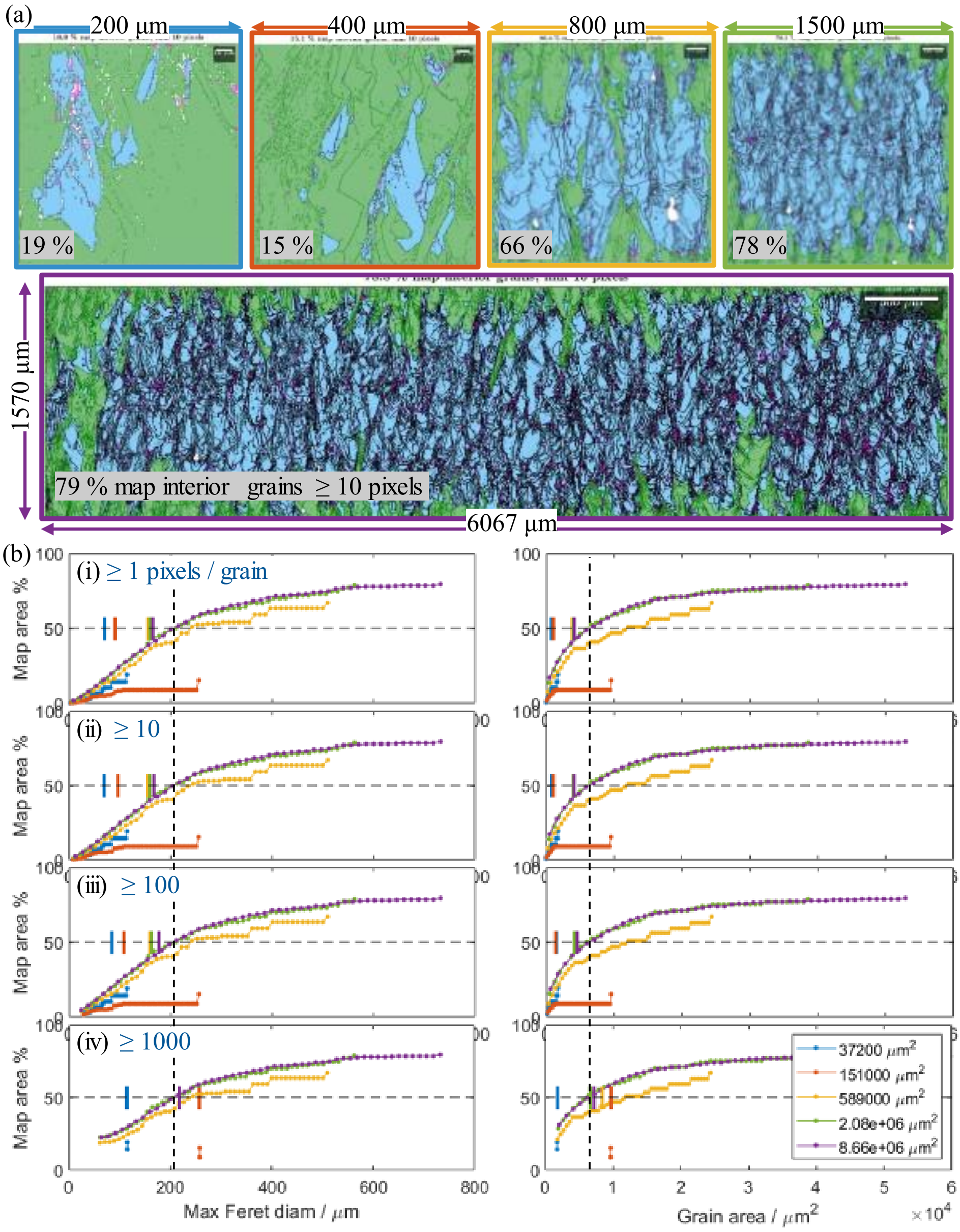

*Figure 3: Effect of grain exclusion on the median grain size. (a) Square map subsets 200² – 1500² µm² cropped from the same map (purple box). The EBSD grain maps show map edge grains in green, small grains < 10 points in pink, and map interior grains in blue. The grey-highlighted text is the area fraction of map interior big grains which may be used for grain size computation according to ISO 13067. (b) Cumulative histograms weighted by indexed map area fraction, for the five EBSD maps in (a), showing MFD (left column) and grain area (right column), and minimum grain size thresholds from 1 to 1000 points (rows). The indexed-area-weighted median grain sizes are marked by dashed black vertical lines. Conventionally area-weighted median grain sizes for each subset are plotted as vertical | markers at the 50 % map area*

Grains can be excluded from analysis if they are too small, or if they cut the map edge. The size of individual small grains has higher uncertainty than larger grains, and (misindexed) noise points may be mistaken for small grains, so ISO 13067 and ASTM E2627 specify that small grains below a threshold size (10 and 100 points respectively) should be excluded from the analysis. The size of grains which cut the map edge cannot be measured directly, and both standards specify that they should be excluded from grain size analysis.

Figure 3a shows square subsets of different sizes cropped from the same large map (purple rectangle). When the map field of view is small with respect to the grain size, the majority of the map area is excluded as edge grains (green grains). The area-weighted grain size is underestimated for small image fields, because large grains are statistically more likely to intersect the map edges and be excluded.

The Miles-Lantejoul correction, recommended in ISO 13067, recognises that the probability that a grain cuts a map edge is equal to its length fraction perpendicular to that edge, and the grain size distribution may be weighted to correct the bias towards excluding larger grains [14]. However, this is not trivial to implement practically in AM microstructures, where a wide range of grain shapes and aspect ratios can contribute to each histogram bin: for example, a long and thin grain is more likely to cut the map edge than an equiaxed grain with similar grain area.

A better area-weighting method is shown in the cumulative histograms in Figure 3b. Here, grain size is weighted by its area fraction of the indexed EBSD map area. If small grains below a threshold size are excluded from the analysis, their area fraction is appended to the small-grain end of the cumulative histogram.

For example, in the 800 × 800 $\mu m^2$ map (Figure 3a, yellow box), 7.7 % of EBSD map points were not indexed. Out of the indexed points, 0.6 % belong to grains < 10 points, 66 % belong to large interior grains, and 33 % belong to large grains that intersect the map edge. Therefore, the bottom 0.6 % and top 33 % of the corresponding cumulative histograms in (Figure 3b-ii, yellow plots) are left empty, to account for the excluded small grains and map edge grains respectively. If grain size values are read from a cumulative histogram, the grain size uncertainty component from histogram bin precision should be modelled as a Type B uncertainty with rectangular distribution.

The variation in indexed-area-weighted median is ± 1.5 % half-range for MFD (207 µm), and ± 9.3 % for grain area (6290 $\mu m^2$). In contrast, conventional area-weighted medians have much higher uncertainties (179 µm ± 16 % for MFD, and 4875 $\mu m^2$ ± 33 % for grain area), since their values change significantly with image field size and minimum grain size threshold (Figure 3b, coloured | markers).

## 2.5 Microstructure sampling

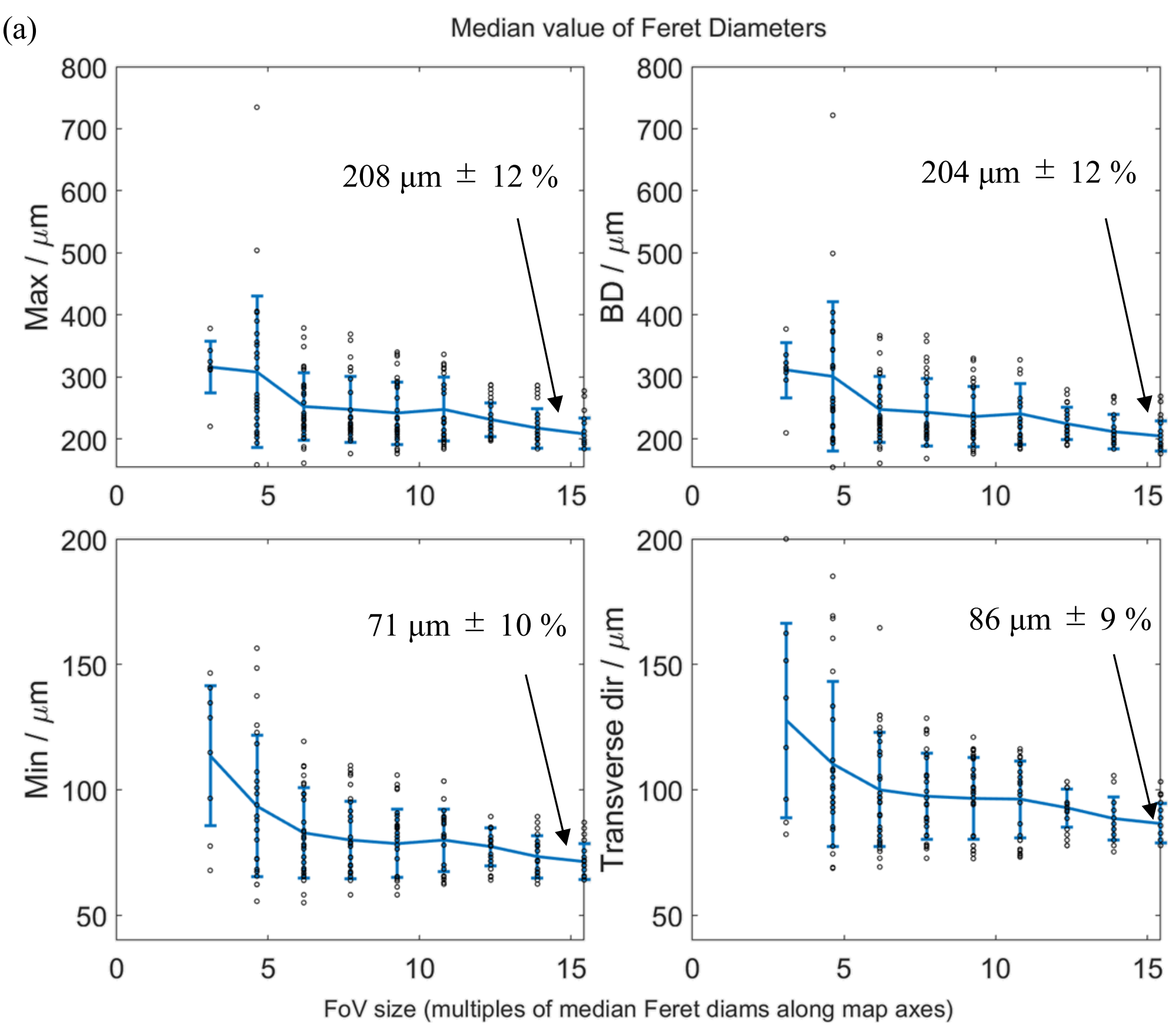

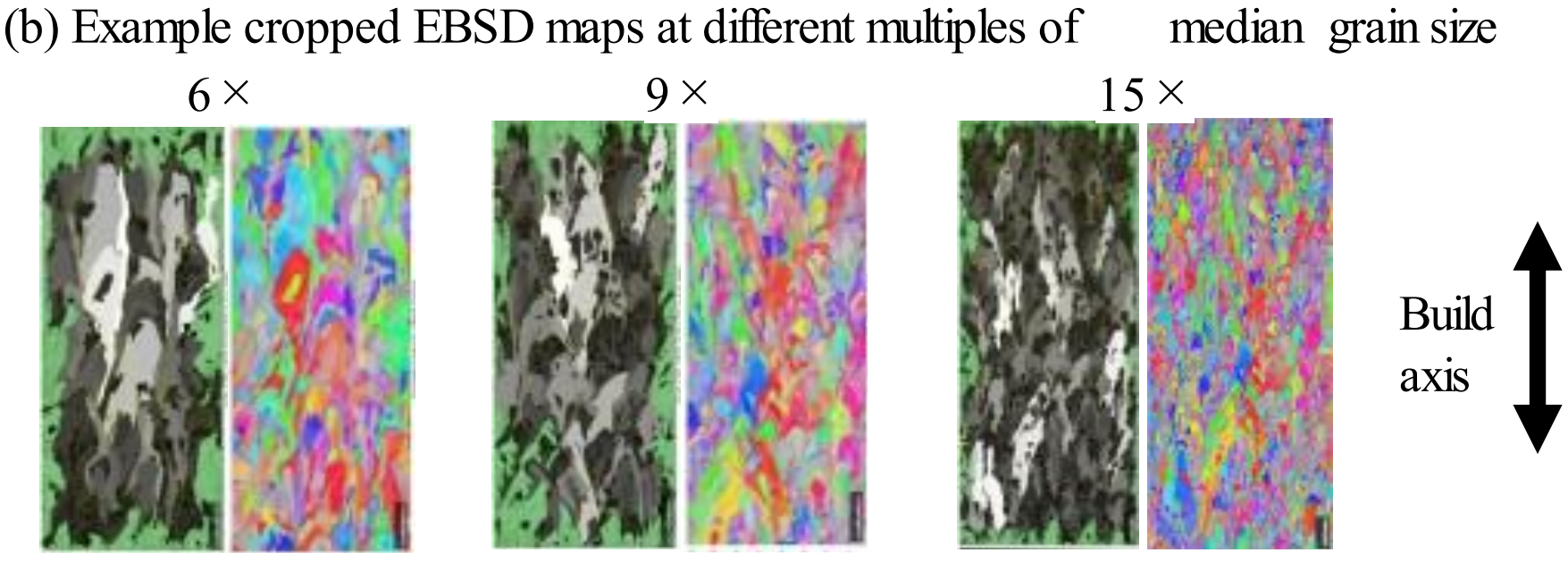

*Figure 4: EBSD map sampling uncertainty components calculated from bootstrap analysis. (a) Indexed-area-weighted medians of Feret diameters along different directions. Scatter points show grain sizes from each individual map, line plots show the mean grain size from 30 samples, and error bars show 1 standard deviation of the measured grain size. (b) Example EBSD grain maps (left) and IPF orientation maps (right) at different FoV sizes. FoV size is expressed as a multiple of the indexed-area-weighted median Feret diameter along the map axes.*

An appropriate field of view size should be a representative sample of the microstructure, as well as contain enough grains to build up the cumulative histogram distributions in Figure 3. Figure 4 shows the sampling uncertainty component to the measured average grain size for different EBSD map field of view (FoV) sizes, using the bootstrap analysis described in Reference [11].

30 EBSD map subsets per FoV size were cropped from a larger EBSD map for grain size analysis. The cropped FoV aspect ratio was selected to follow the ratio of the indexed-area-weighted median Feret diameters along the EBSD map axes (202 µm and 86 µm along and transverse to the build direction respectively). The FoV sizes (horizontal axis in Figure 4(a)) are scaled as multiples of these median Feret diameters, and Figure 4(b) show qualitative examples of grain size and EBSD IPF orientation maps at a range of FoV sizes.

Four indexed-area-weighted median Feret diameters were calculated from each subset: the maximum and minimum Feret diameters, and the Feret diameters along the EBSD map directions parallel and transverse to the AM build direction. The MFD and Feret-BD diameters are similar, but the minimum Feret diameters are significantly smaller than the Feret-Transverse diameters because the grain shapes are far from elliptical.

All four grain size plots in Figure 4(a) follow a similar trend: Sampling uncertainty (error bar widths) increases with decreasing FoV size, but he mean of the average grain sizes is stable for FoV sizes > 7 ×. As the FoV becomes smaller, the mean value increases because of large outliers (black scatter points) where the map FoV is dominated by one large grain.

The sampling uncertainty is around 10 % for FoV sizes of 15 × grain lengths. This a lower-bound estimate of the random sampling uncertainty because the sampled areas overlap when bootstrapping, which is significant for larger FoV sizes.

## 2.6 EBSD map step size and grain boundary threshold angle

Grain size is intrinsically linked to the EBSD map step size and boundary threshold angle in typical AM microstructures, which have large intragranular rotations and sub-grain boundary structures. The distinction between grain boundaries and sub-grain boundaries is somewhat arbitrary and depends on the grain boundary threshold angle value, as well as the grain reconstruction method. The EBSD map step size affects the effective grain boundary threshold angle during grain reconstruction, because grain boundaries are defined by the misorientation angle between adjacent points. When the step size is extremely large, grains can also be broken up because pixel-connectivity is lost in the thin regions of non-convex grains.

A longer step size decreases the measured grain size, because grain boundaries with misorientation just above the threshold angle at a larger step size are no longer classified as grain boundaries at a smaller step size, since the same misorientation is split between more map points. Figure 5 shows an example of this: the EBSD map (a) was 2 × 2 binned to double its step size in (b). The maximum grain size becomes smaller when the step size is Figure 5 doubled, and the circled grains highlight a few single grains at a smaller step size which have been split into multiple grains at a larger step size.

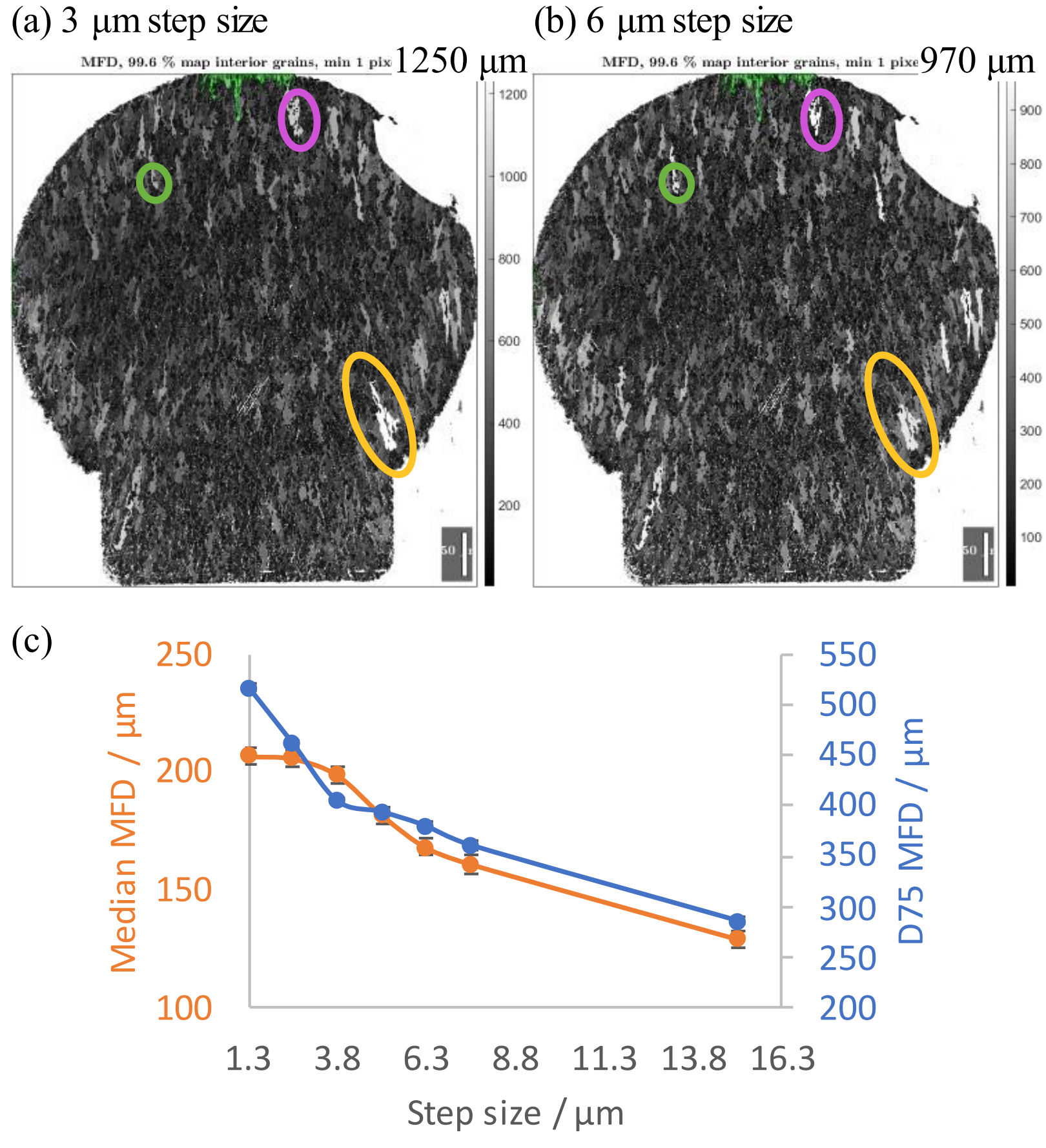

*Figure 5: Effect of step size on grain size. (a) Original and (b) 2 × 2 binned EBSD maps of Hastelloy-X with MFD colour scaling, where black is the smallest grain and white is the largest grain within each map. (c) Systematic decrease of median (orange) and 75th percentile (blue) MFD with increasing step size, obtained by binning the EBSD map in Figure 3(a) from 1.25 to 15 µm.*

# 3 Application to different microstructures

## 3.1 Hastelloy-X

Three AM parts were made by selective laser melting using the same powder and identical build parameters except for the laser power variation mode (square, sinusoidal, triangular waveform). EBSD maps were acquired from four different SEMs using a wide range of experimental parameters, and spatially binned to match the step sizes to as close as possible to 2.5 µm. All maps had at least 90 % indexing and 11 grains along the build direction. A 10° boundary threshold angle was used throughout.

The uncertainty component from grain clean-up was calculated from the difference between no clean-up and maximum possible clean-up, and treated as a Type B uncertainty with rectangular distribution. The uncertainty component from grain size histogram binning was the histogram bin half-width, and treated as a Type B uncertainty with rectangular distribution. The uncertainty component from microstructure sampling was calculated from 30 bootstrap subsamples as described in Section 2.5 while maintaining the same map length along the build direction, and treated as a Type A uncertainty. These were converted to standard uncertainties, combined by adding in quadrature, and finally expressed as an expanded uncertainty with coverage factor $k = 2$.

Table 2 shows the average grain sizes of these three parts, expressed as median indexed-area-weighted MFDs, and as well as another Ni-base superalloy from an unrelated AM build for comparison. Figure 6 shows grain size maps and cumulative histograms of the grain size distributions of the same maps. The average grain size and grain size distributions of the three similar AM parts are similar, whereas unrelated Ni sample has a very different grain size and size distribution. Maps which are shorter along the build axis have more stochasticity and tend to overestimate the grain size at higher quantile values, which is consistent with the bootstrap analysis results in Figure 4.

Any potential effect of laser power variation mode (triangle, sinusoidal or square) on grain size distribution is smaller than the grain size measurement uncertainty.

*Table 2: Average grain sizes and uncertainties for four EBSD maps of Hastelloy-X and one unrelated AM Ni sample. *For the Triangle dataset, cropped map at full length was not much smaller than the full map, so the map length was halved for the sampling uncertainty calculation to avoid too much overlap in the bootstrap model.*

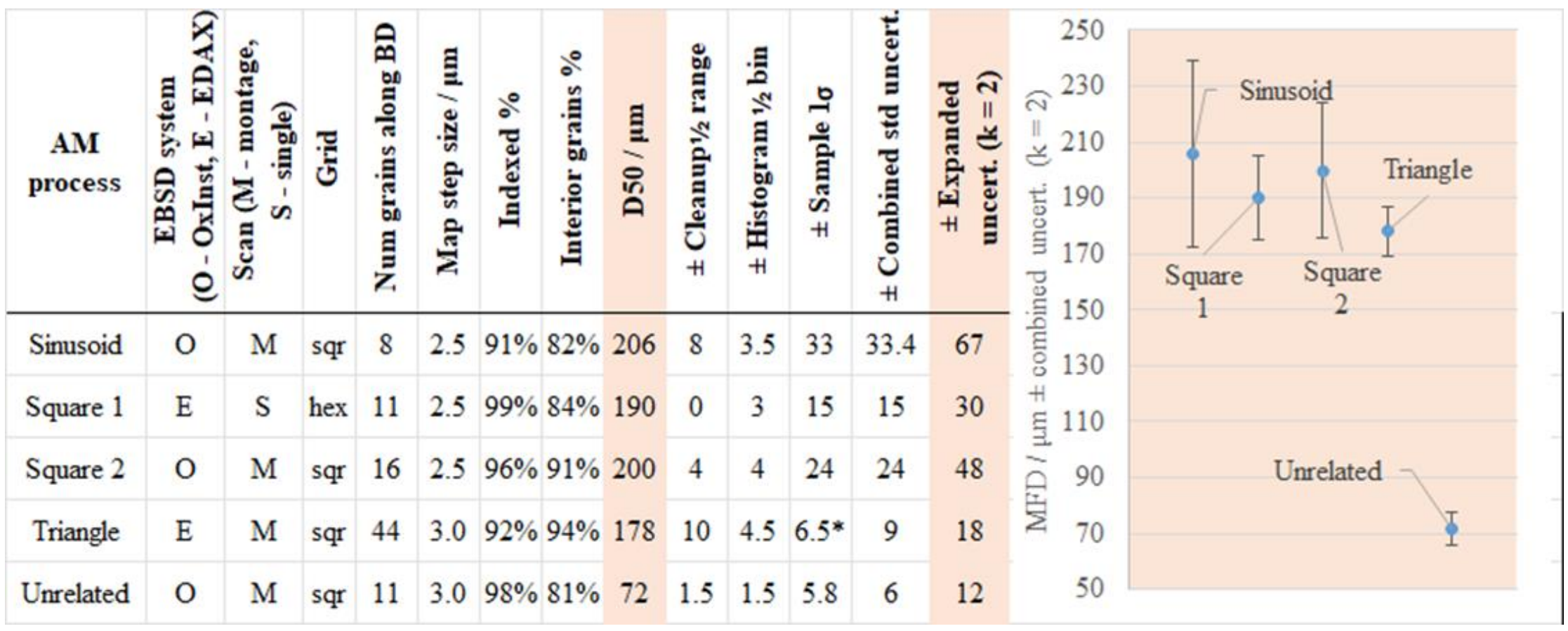

| AM process | EBSD system (O - OxInst, E - EDAX) | Scan (M - montage, S - single) | Grid | Num grains along BD | Map step size / µm | Indexed % | Interior grains % | D50 / µm | ± Cleanup½ range | ± Histogram ½ bin | ± Sample 1σ | ± Combined std uncert. | ± Expanded uncert. (k = 2) |
|---|---|---|---|---|---|---|---|---|---|---|---|---|---|
| Sinusoid | O | M | sqr | 8 | 2.5 | 91% | 82% | 206 | 8 | 3.5 | 33 | 33.4 | 67 |
| Square 1 | E | S | hex | 11 | 2.5 | 99% | 84% | 190 | 0 | 3 | 15 | 15 | 30 |
| Square 2 | O | M | sqr | 16 | 2.5 | 96% | 91% | 200 | 4 | 4 | 24 | 24 | 48 |
| Triangle | E | M | sqr | 44 | 3.0 | 92% | 94% | 178 | 10 | 4.5 | 6.5* | 9 | 18 |
| Unrelated | O | M | sqr | 11 | 3.0 | 98% | 81% | 72 | 1.5 | 1.5 | 5.8 | 6 | 12 |

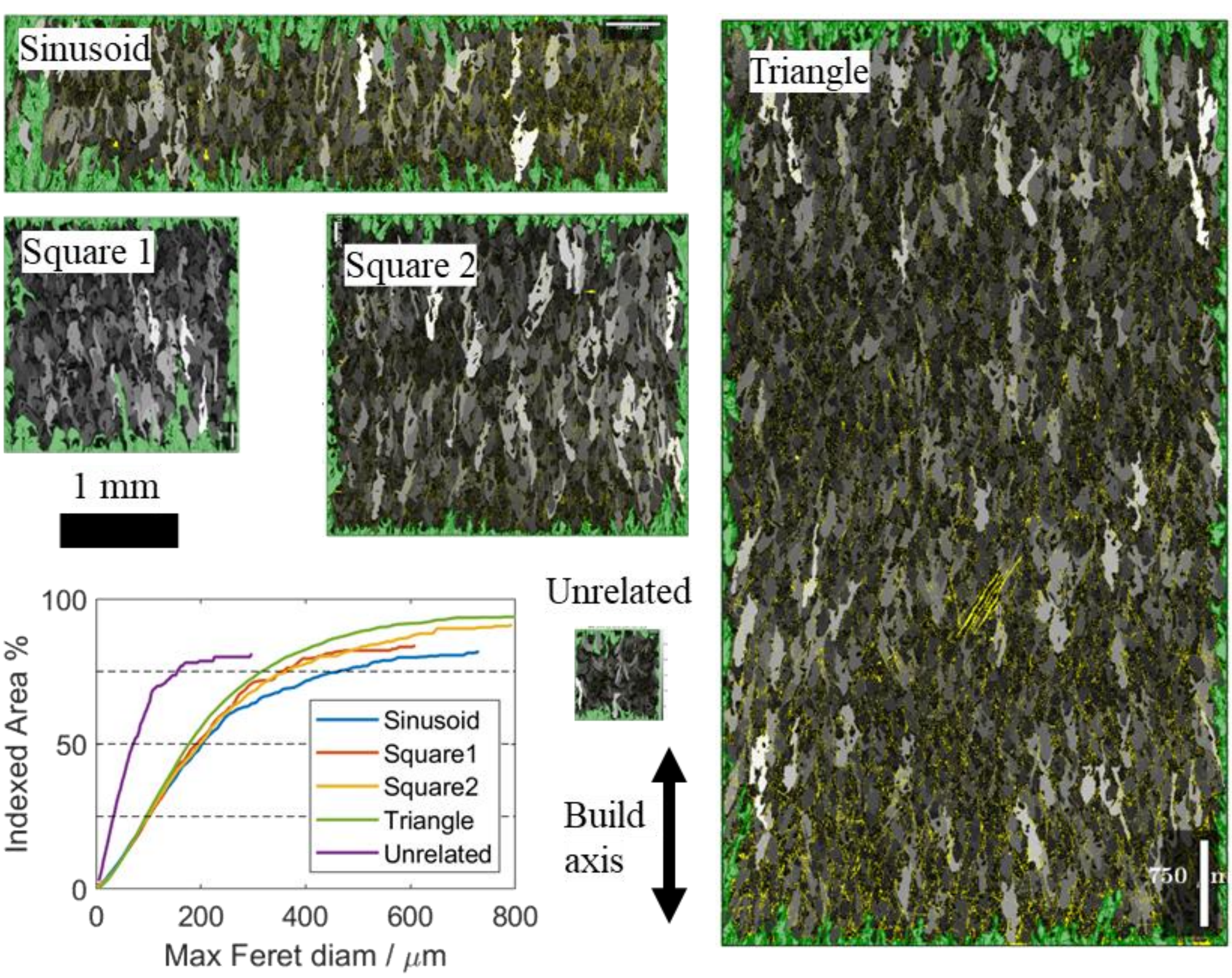

*Figure 6: Grain maps and grain size distributions for the five EBSD maps analysed in Table 2.*

## 3.2 AlSi10Mg

### 3.2.1 Large components

Two different AM builds of AlSi10Mg were analysed in this section. Both contained pores and a banded microstructure along AM build layers. Grains oriented with [001] parallel to the build direction (red in the orientation maps) are larger and elongated approximately along the build direction, whereas grains of other orientations are smaller and equiaxed. Figure 7 and Figure 8b show AlSi10Mg EBSD maps from two unrelated AM builds but similarly large components.

The EBSD maps in Figure 7 were acquired with 20 kV electrons, on two orthogonal faces with build directions perpendicular and in-plane (Figure 7a and b respectively). The indexing rate for these maps were low (65 % for Figure 7a and 81 % for Figure 7b) because of sample porosity (e.g. white arrow in Figure 7a) and inadequate EBSD spatial resolution at 20 kV. The unindexed points were not randomly positioned, but disproportionately affected areas with smaller grains. These maps are not compliant with any existing standards.

In contrast, the AlSi10Mg EBSD map in Figure 8(b) contained fewer pores, and the EBSD map was acquired at 15 kV to improve spatial resolution, resulting in a 95 % indexing rate.

The uncertainty components due to grain clean-up were ± 13 % for the map in Figure 7a (65 % indexed) and ± 4 % for the map in Figure 8b (95 % indexed). These relative uncertainty components from grain clean-up are consistent with the results from Hastelloy-X described in Table 2.

More complex grain properties, such as the banded grain structure and correlations between preferred orientation and grain size or aspect ratio, cannot be captured by the median MFD statistic. However, the median MFD is able to show the differences between the two AM builds, as well as the elongation along the build direction.

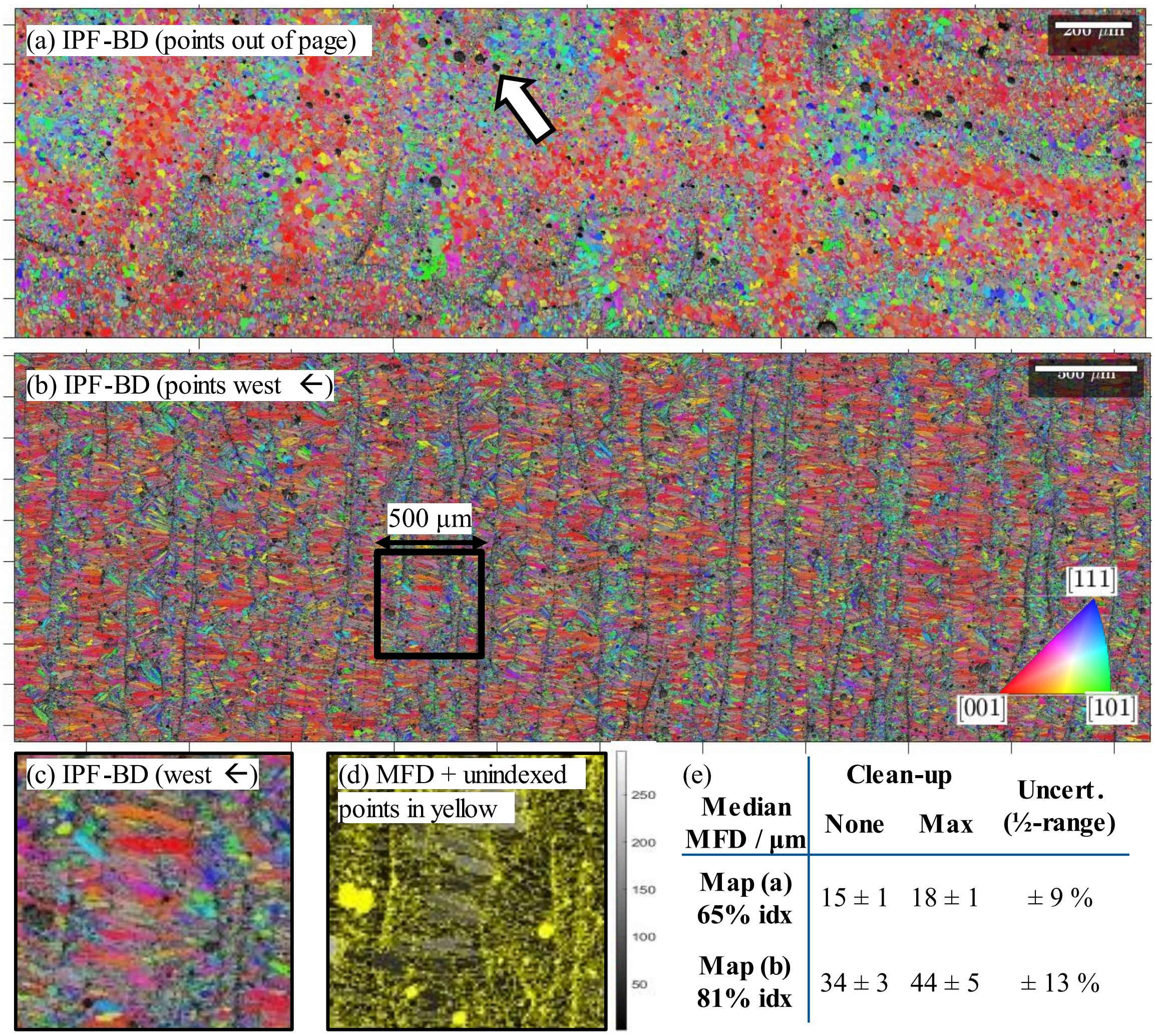

| Median MFD / µm | Clean-up None | Clean-up Max | Uncert. (½-range) |
|---|---|---|---|
| **Map (a) 65% idx** | 15 ± 1 | 18 ± 1 | ± 9 % |
| **Map (b) 81% idx** | 34 ± 3 | 44 ± 5 | ± 13 % |

*Figure 7: AlSi10Mg EBSD maps (20 kV, 65 – 81 % indexed). (a) EBSD pattern quality and orientation map showing IPF directions along the build direction, which points out of page. White arrow points to an example pore. (b) EBSD pattern quality and orientation map showing IPF directions along the build direction, which points west (left) in the map. (c,d) Enlarged portions of the 500 µm wide box in (b) showing unindexed points concentrated at pores and near small grains. (e) Median MFDs and grain size uncertainty due to grain clean-up. The MFD ± values are the cumulative histogram bin widths used for MFD computation.*

### 3.2.2 Lattice structures

Figure 8a shows AlSi10Mg EBSD maps of lattice strut cross-sections where the AM build parameters are identical to those of Figure 8b but the component size is small and has a complex geometry. The grain elongation direction no longer follows the build direction but instead depends on the local heat flow environment, which is related to the locations of neighbour struts, as well as the distance from the solid plate they were built up from. In the third layer up from the solid plate in Figure 8a, the microstructure switches from columnar to equiaxed and dendritic. Similar microstructural variations have been observed in AlSi10Mg subjected to different heat treatments [15].

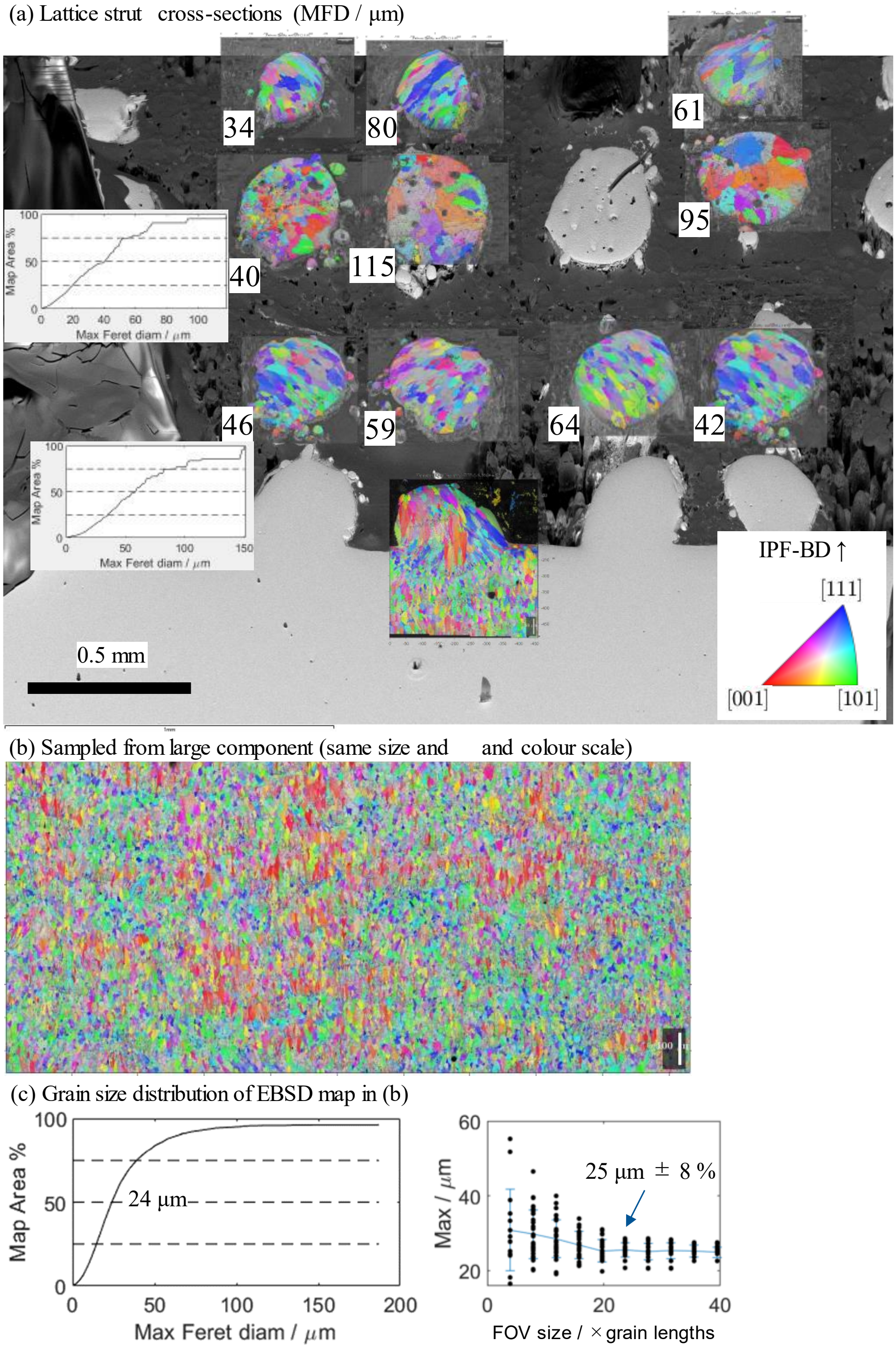

*Figure 8: AlSi10Mg EBSD maps from (a) lattice struts and (b) a large solid part. Text annotations are the median MFD for each strut in µm, and cumulative histogram plots for two example strut sections. (c) Grain size cumulative histogram and FOV size calibration plot calculated from the EBSD map in (b).*

# 4 Discussion

## 4.1 Sample selection

The microstructure of an AM material depends on the shape of the built component, because the local cooling and reheating conditions, which are related to the component shape, can change the microstructure.

This is very obvious in the strut cross-sections in the AlSi10Mg lattice structures (Section 3.2.2), which have completely different microstructures from the (nominally) larger parts of the same AM build, and which changed with the three-dimensional geometry of the surroundings. The grain size and any microstructural features in these regions apply only to regions with the same local geometry, including distance from nearby thermal masses. Since each strut section contains too few grains to build up a grain size distribution, multiple similar samples would be necessary to measure the average grain size from small sections such as this.

Local effects are present even in large components, where grains near a surface or side-wall can be different to internal grains. For example, the surface grains in the Hastelloy-X sample in Figure 5 are elongated along the radial direction, whereas internal grains are elongated along the build direction (this is explored further in Reference [12]). In the interlaboratory comparison study in Reference [1], a major uncertainty component was related to whether or not participants chose to crop out fine grains near the AM build surface from the analysis FoV.

## 4.2 EBSD analysis parameters

Section 2.6 showed that grain sizes measured from EBSD maps acquired at different step sizes are not directly comparable, even if the boundary threshold angles are the same, because the high temperature gradients of AM processes can generate microstructures with sub-grain boundaries and large intragranular misorientations. This is different to microstructures with fully recrystallised grains separated by high-angle grain boundaries, where EBSD map step size and grain boundary threshold angle can be treated as independent parameters. Therefore, grain size should always be reported with its EBSD map step size and grain boundary threshold angle in such microstructures.

There is no objectively optimal value for the threshold angle, but it affects the measured grain size greatly, and different values can be appropriate for different material types. Identical threshold angles (and step sizes) must be used when comparing EBSD grain sizes.

Since AM microstructures have very wide size distributions, a good grain size spatial resolution should be balanced with a sufficiently large FoV. The guidance on EBSD map step size selection in ISO 13067 (around 10 % of the average grain diameter, in this case minimum Feret diameter) is a good starting point. Pixellation uncertainties in reconstructing very small individual grains are not important for measurement of average grain size, as it only affects the small-grains tail of the grain size cumulative histogram, not the median (Figure 3b). However, AM grain shapes are often non-convex so using a too-large step size can break up thin sections of large grains. 20 % of the average grain diameter, recommended in Reference [7], may be appropriate for convex polygonal grain structures, but is too coarse for resolving AM structures.

The FoV size should be as large as practical and follow the aspect ratio of the grains, since a small FoV increases the random sampling uncertainty (Section 2.5) which dominates the combined uncertainty (Table 2). 20 grain lengths per FoV were required to achieve a sampling uncertainty of about 10 %. A lower bound for the sampling uncertainty contribution was approximated using a bootstrap model for this work, and should be determined by sampling non-overlapping fields of view in a real measurement.

A second effect of a too-small FoV is that a larger proportion grains are grains will cut the map edge and be excluded (Section 2.4). In a small FoV, few or no grains contribute to the large-grained end of the indexed-area-weighted cumulative histogram, so the median grain size has higher uncertainty or even an undefined value (if more than 50 % of the map has been excluded).

## 4.3 Representing average grain size

The MFD is a suitable grain size metric for representing sectional grain size from an EBSD map in AM microstructures. Figure 4 showed that the MFD is a good approximation of Feret diameter along the build direction, since AM grains tend to be elongated along the build direction. The minimum Feret diameter is different to the Feret diameter perpendicular to the build direction, but either one may be used as a measure of grain shape anisotropy.

The indexed-area-weighted median MFD and cumulative histogram plot, described in Section 2.4, is a suitable summary statistic for representing the average. It offers several advantages over conventional histograms and (arithmetic or geometric, area- or number-weighted) mean grain size:

Firstly, the median grain size can be read directly from the 50 % point of the cumulative histogram. Cumulative histograms are relatively insensitive to histogram bin width, which can change the visual appearance of a conventional histogram plot even if the underlying data is identical, and the median does not assume anything about the theoretical distribution shape.

Secondly, the grain size distribution is insensitive to minimum grain size threshold, because excluding small grains does not shift, but only truncates, the distribution tail at the small grain end. This is shown in the histograms of Figure 3b i-iv, with minimum grain size thresholds of 1 to 1000 points.

Thirdly, a too-small field of view is obvious from the cumulative histogram only, without having to inspect the EBSD maps separately. For example, you can tell that the three smallest subsets in Figure 3a are too small from the shape of the histograms in Figure 3b: the $(200\ \mu m)^2$ (blue) and $(400\ \mu m)^2$ (red) subsets never reach 50 %, so the median value is not available at all; the $(800\ \mu m)^2$ (yellow) histogram reaches 50 %, but the appearance of large steps at $\geq 4000\ \mu m^2$ grain area ($\geq 35$ % map area) indicates that very few grains contribute to the histogram distribution in this size range, and the median value is highly uncertain.

Two simplifications are implicit in the indexed-area-weighting method: firstly, that grains cut by the edge of the map are the biggest grains (and therefore contribute only to the large-grain end of the cumulative histogram); secondly, that unindexed EBSD map points are spread uniformly over the histogram, i.e. there is no correlation between grain size and indexing rate. Both of these are simplifications – it is possible for a small grain to be cut by the edge of the map, and the unindexed points are concentrated at grain boundaries, so the size of grains with shorter perimeters are underestimated. The effect of unindexed EBSD map points can be minimised by ensuring a high (> 90 %) indexing rate.

### 4.4 AM grain size measurement procedure

Recommendations for grain size measurement, determined from the analyses in Sections 2 and 3, are summarised in the list below.

(1) **Average grain size shall be reported with grain boundary threshold angle and EBSD map step size** whenever significant sub-grain structure or intragranular misorientations are present. Grain reconstruction should be performed with a single g.b. threshold angle value.

(2) The **indexed-area-weighted median** shall be reported as the primary grain size statistic. The 25$^{th}$ and 75$^{th}$ percentiles, or other quantile values, may also be reported.

(3) Local dimensions of the AM component shall be reported if it has a **complex geometry that affects the microstructure** compared to equivalent bulk material.

(4) **EBSD map clean-up may be performed** with up to 5 % change to map orientations. The grain size uncertainty component from clean-up should be measured by repeating the analysis without clean-up and with the maximum level of clean-up, and modelled as a Type B uncertainty with rectangular distribution. The EBSD map clean-up shall not artificially fill in physical pores or voids.

(5) **Edge grains shall be excluded** from the grain size distribution.

(6) **Small grains may be excluded** from the grain size distribution. If small grains are excluded, their indexed area fraction shall be added to the small-grain end of the cumulative histogram.

(7) The grain size distribution should be **visualised as a cumulative histogram**, weighted by its area fraction of the total indexed area of the EBSD map **('indexed-area-fraction-weighted' cumulative histogram distribution**). Note that the indexed area includes all indexed EBSD map points, including edge grains and small grains.

(8) The grain size distribution should be **visualised as an EBSD map with grain size colour scaling**. The colour scale should be perceptually linear.

(9) The EBSD field of view shall be large enough so that the grain area cumulative histogram does not have discrete jumps at 50 %, which would indicate that few grains are contributing to the median value. As a rule of thumb, the field of view should be at least 20 × larger than the median grain lengths along that dimension. Montage mapping may be used to generate EBSD maps with large fields of view, but should be cropped to a rectangle so that map edge grains can be defined and excluded.

# 5 Summary

The indexed-area-fraction weighted median MFD is a robust metric for representing average grain size from EBSD maps of AM components. A method for measuring average grain size, visualising the grain size distribution, and determining uncertainty components from random sampling, EBSD map clean-up, and cumulative histogram binning precision has been tested on several AM Ni and Al components. A relative combined expanded uncertainty (k = 2) of 20 % can be achieved using this procedure. The total uncertainty is dominated by the random sampling uncertainty component. When measuring AM components with complex part geometries, several similar samples may be required to obtain a statistically significant sample.

# 6 Acknowledgements

Steven Ravenscroft, Linda Orkney, Helen Jones, Tony Fry, Maria Lodeiro and Peter Woolliams from the Advanced Engineering Materials group at NPL contributed to the related project and provided samples. Sethupathi Rangaraj from Manchester University contributed background information for interpreting the AlSi10Mg microstructures.